\documentclass[floats,aps,twocolumn,prl,psfig,graphicx]{revtex}
\usepackage{graphicx}
\usepackage{amsmath}

\begin{document}
\title{Sequestration of noble gases in giant planet interiors}
\author{Hugh F. Wilson$^1$ and Burkhard Militzer$^{1,2}$} \affiliation{Departments of Earth
  and Planetary Science$^1$ and Astronomy$^2$, University of California Berkeley}

\author{The Galileo probe showed that Jupiter's atmosphere is severely depleted in neon compared to protosolar values. We show, via ab initio simulations of the partitioning of neon between hydrogen and helium phases, that the observed depletion can be explained by the sequestration of neon into helium-rich droplets within the postulated hydrogen-helium immiscibility layer of the planet’s interior. We also demonstrate that this mechanism will not affect argon, explaining the observed lack of depletion of this gas. This provides strong indirect evidence for hydrogen-helium immiscibility in Jupiter.}
\maketitle

Jupiter is the most extensively probed and best understood of the
giant planets, but many questions regarding its detailed composition,
formation, and interior structure remain unanswered. One issue of
major importance to structural models is the question of whether
hydrogen and helium mix homogeneously throughout the planet or whether
a layer of hydrogen-helium immiscibility exists deep within the
interior
\cite{salpeter-astropj-73,stevenson-prb-75,stevenson-astropj-77-i}.
In the immiscibility layer, helium would form dense droplets which
would rain down into the deeper interior and redissolve, resulting in a gradual and
ongoing transfer of helium from regions above the immiscibility layer
to regions below. Such a layer almost certainly exists in Saturn, as
evident from the observed depletion of helium from its upper
atmosphere (compared to protosolar values) and the apparent excess
luminosity of the planet \cite{stevenson-science-80}. For the hotter interior of
Jupiter the case is less clear since there is no measurable excess
luminosity and the observed helium depletion from the upper atmosphere
is quite small (0.234 by mass compared to 0.274 in the protosolar
nebula \cite{niemann-science-96,vonzahn-jgr-98,lodders}). Theoretical
attempts to determine the pressure/temperature range in which H and He
are immiscible using successively more sophisticated levels of theory
\cite{stevenson-astropj-77-i,stevenson-astropj-77-ii,straus-prb-77,
  hubbard-astropj-85,klepeis-science-91,pfaffenzeller-prl-95,lorenzen-prl-09,morales-pnas-09}
have produced quite different results, however recent work
\cite{morales-pnas-09,lorenzen-prl-09} provides a hydrogen-helium
immiscibility line which is very close to the Jupiter isentrope in the
100--300 GPa region.

\begin{figure}[htbl]
\centerline{
 \includegraphics[width=4.5cm]{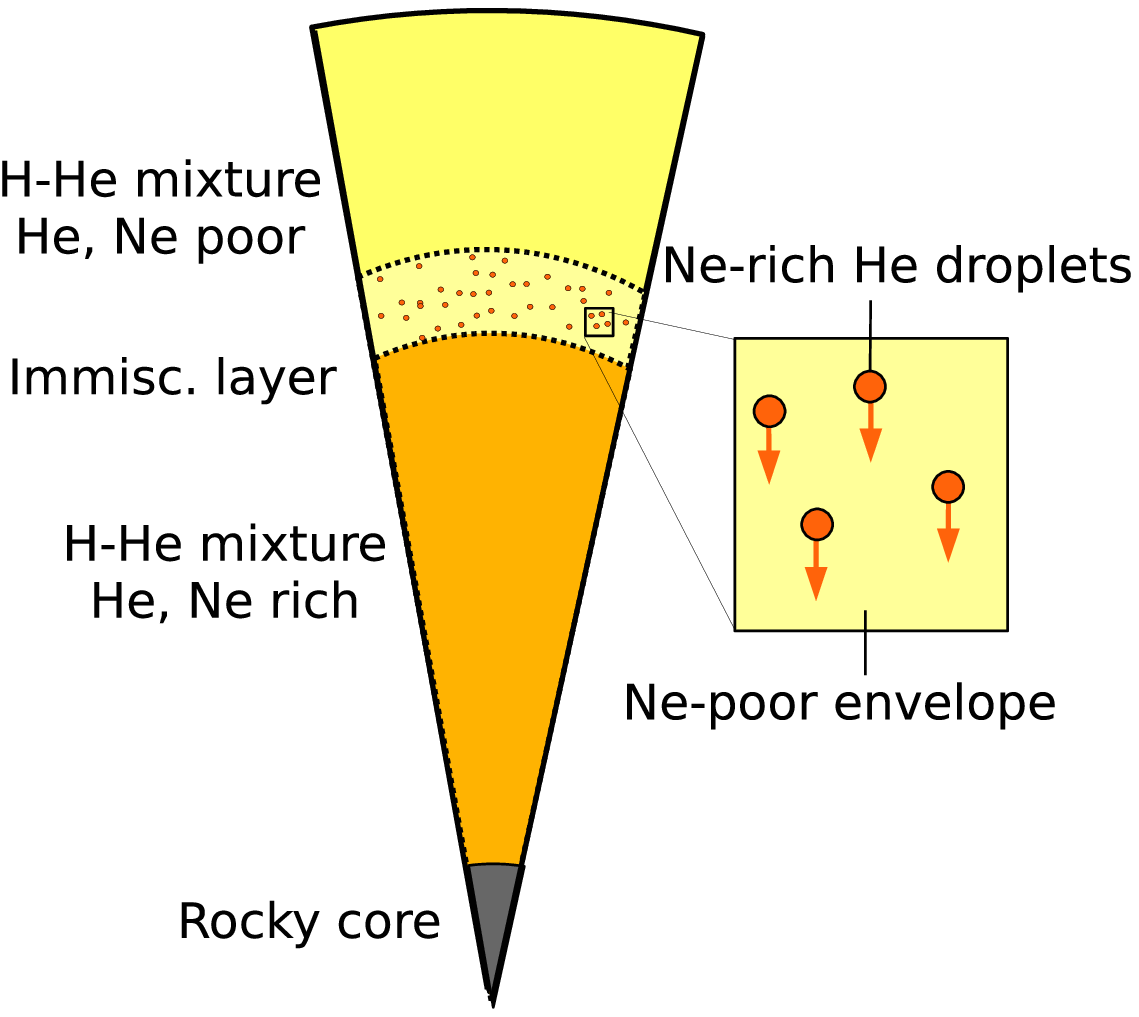}
 \includegraphics[height=4.5cm,angle=90]{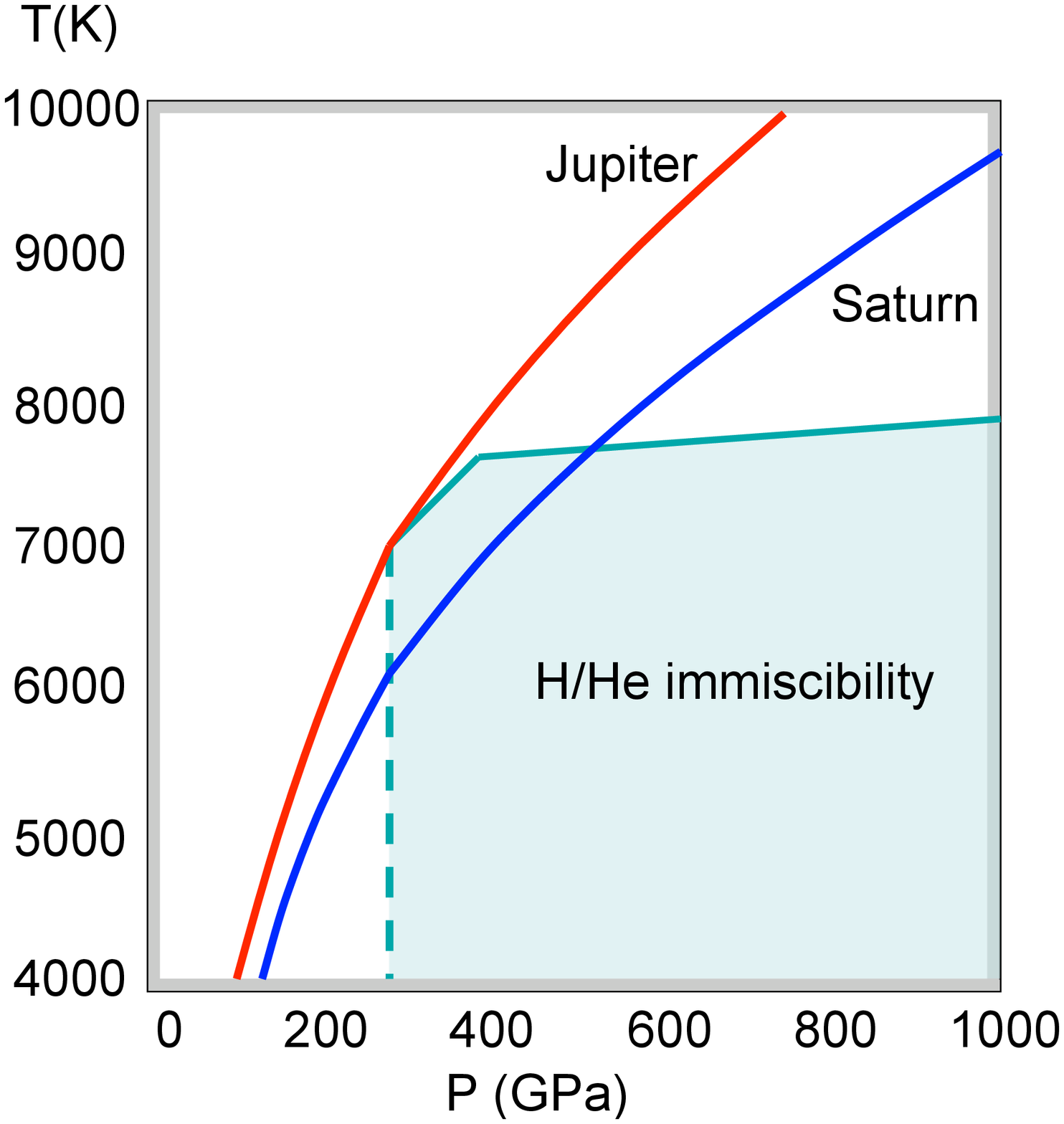}
}
\caption{ (left) Schematic depiction of the
interior of a gas giant (e.g. Jupiter or Saturn) with a layer of H-He
immiscibility. Helium-rich droplets form within the immiscibility
layer and rain downwards, leading to a slow increase in the helium
concentration in the deep interior. Neon is absorbed within the
droplets and carried out of the upper atmosphere.(right) P/T curves for Jupiter and Saturn combined with the
  location of the H-He immiscibility region determined in the work of
  Morales \emph{et al} \cite{morales-pnas-09} assuming an overall He
  atomic molar concentration of 0.0847.}
\label{schematic}
\end{figure}

The strongest evidence for H-He immiscibility may in fact come from
the depletion of neon.  Jupiter's upper atmosphere was found by the
Galileo entry probe to be extremely deficient in neon: although neon
makes up $1/600$ by mass in the solar system it comprises only
$1/6000$ by mass in Jupiter's upper atmosphere
\cite{niemann-science-96}. Prior to these measurements, Roulston and
Stevenson \cite{roulston-eos-95} proposed that hydrogen-helium
immiscibility could lead to neon depletion on the assumption that neon
would preferentially dissolve in the helium-rich phase in the
immiscibility layer. This would lead to Jupiter's neon content being
gradually carried down within the helium droplets and concentrating in
the deep interior. When hot fluids enriched in He and Ne are convected
upwards they are subjected to differentiation again. There is,
however, a lack of direct experimental evidence for whether Ne will
indeed preferentially dissolve in the helium phase as proposed. The
pressure-temperature conditions corresponding to phase separation can
currently only be attained in shock-wave experiments lasting only tens
of nanoseconds \cite{knudson-prl-03} and which are hence ill-suited to
study questions of phase separation and partitioning. Is it also not
known why the chemically similar noble gas argon is not seen to be
depleted but instead is present at slightly above-solar concentrations
($\approx$ 1.6 times solar \cite{niemann-science-96}) comparable to
most other detectable trace heavy elements in Jupiter, and whether
this indicates that the depletion mechanism acting upon neon does not
act upon argon or whether it indicates a very high initial argon
concentration. In order to resolve these issues, here we present
\emph{ab initio} free energy calculations with a view to determining
the solubility behaviour of neon and argon in H and He phases at
pressures corresponding to the postulated immiscibility region.

The distribution of a trace species such as Ne or Ar between
coexisting phases is dependent upon the Gibbs free energy of transfer,
$\Delta G_{\text{Tr}}$, being the change in $G$ when an atom of the
trace species is moved from one phase to the other at constant $P$ and
$T$, in this case

\begin{equation}
\Delta G_{\text{Tr}} = G({\text{He + Ne}}) + G({\text{H}}) - G({\text{H + Ne}}) - G({\text{He}}).
\end{equation}

Here we compute $\Delta G_{\text{Tr}}$ for neon and argon in pure H
and He within the density functional theory molecular dynamics
(DFT-MD) framework, within a temperature and pressure range of
100--300 GPa and 3000 to 7000~K. Determination of free energies from
MD is a nontrivial problem for which a number of
methods have been developed.  We use a two-step coupling constant
integration (CCI) approach \cite{alfe-nature-00} similar to that
recently applied by Morales \emph{et al.}  \cite{morales-pnas-09}, in
which the Gibbs free energy of the DFT system is determined by
adiabatically transforming the system in two steps: (a) from the DFT
system to a system of atoms governed by a classical potential and (b)
from the classical system to a noninteracting gas.

The CCI method provides a general scheme for computing the Helmholtz
free energy difference between two systems governed by potential
energy functions $U_1$ and $U_2$. Constructing artificial system with
potential $U_\lambda = (1-\lambda)U_1 + \lambda U_2$, the free energy
of system 2 may be expressed as:

\begin{equation}
F_2 = F_1 + \int_0^1 d \lambda \langle U_2 - U_1 \rangle_\lambda,
\end{equation}

where at each integration point, the average is taken over a sample of
configurations obtained in the $U_\lambda$ system. Since the difference in Gibbs free
energy between a system at two different pressures can be found by the
thermodynamic integration $G(P_2) - G(P_1) = \int_{P_1}^{P_2} V dP$,
we performed all CCI calculations at pressures close to 200~GPa and
then integrated the equation of state for each system outwards to
obtain $G$ values at pressures from 100 to 300~GPa.

The first part of the CCI was the integration from the noninteracting
system to the classical system.  The classical potential used was a
pair potential of a modified Yukawa form \cite{morales-pnas-09}:

\begin{equation}
U(r) = a \left( \frac{e^{-br}}{r} + \frac{e^{-b\left(L - r \right)}}{\left( L - r \right)} - 4 \frac{ e^{bL/2}}{L} \right),
\end{equation}

for $r < L/2$ and zero otherwise. We set $L = 9.749$ a.u., then fitted
$a$ and $b$ to the $g(r)$ functions of the DFT systems at high
pressure \footnote{The parameters ($a$,$b$) obtained for the pair
  potential between each pair of species were as follows, using atomic
  units: H-H:(1.0,2.5), He-He:(5.5,1.6), H-Ne:(8.0,1.86),
  He-Ne:(8.76,1.34904), H-Ar:(13.0,2.119), He-Ar:(16.5,1.36125)}. The
integration from the noninteracting to classical system used sixteen
$\lambda$ values, spaced more closely in the small-$\lambda$ region
where the computed classical energy varies rapidly.  We checked
potentials obtained by a force-matching method and found that the
numerical results achieved for the free energy of the final system
were not altered by the different potential.

The second part was the integration from the classical forces to the
DFT forces. This was the most time-consuming step and required a
series of DFT-MD runs. We found that the variation in $\langle
U_{\text{DFT}} - U_{\text{classical}}\rangle$ was sufficiently close
to linear in $\lambda$ to allow a fit with only three $\lambda$ points
to be used -- checks against calculations with five $\lambda$ points
resulted in discrepancies smaller than 0.1~eV. All DFT simulations
were performed using the VASP code \cite{vasp}. We used 128 H atoms or
64 He atoms with or without a single Ne or Ar atom inserted per cell.
We used pseudopotentials of the projector-augmented wavefunction type
\cite{paw}, the exchange-correlation functional of Perdew, Burke and
Ernzerhof \cite{pbe}, a cutoff of 1000 eV and eight $k$-points in the
Monkhorst-Pack grid. All MD simulations used a timestep of 0.4~fs and
were run for 5000 timesteps with the first 1000 steps discarded for
equilibration. 

The total Gibbs free energy computed for each system (H,
H+Ne,H+Ar,He,He+Ne,He+Ar) is a sum of five terms:

\begin{eqnarray}
\nonumber G(P_1) = F_{\text {ideal}}(V_0) + \Delta F_{\text {ideal} \rightarrow \text{classical}}(V_0) + \\
\Delta F_{\text{classical} \rightarrow \text{DFT}}(V_0) + P_0V_0 + \int_{P_0}^{P_1} V dP,
\end{eqnarray}

where $P_0$,$V_0$ are the pressure and volume at which the $F$ values
were computed, and $P_1$ is the pressure of interest. The CCI
procedure was undertaken at pressures within 1\% of 200~GPa and $VdP$
integration was used to correct the values back to 200 GPa
exactly. Pressure-volume curves were obtained from a series of five MD
simulations on each system at pressures spaced from 100 to 300 GPa,
and by fitting the resulting data points with a piecewise power law
fit.

In order to validate this method, we also performed simulations via an
alternative free energy calculation method based on the particle
insertion formalism of Widom \cite{widom-jchemphys-63}. Using only the
$\Gamma$ point for Brillouin Zone sampling, we computed the free
energies associated with the insertion of Ne and Ar into He and H
cells at volumes corresponding to a Wigner-Seitz radius for the
electrons of 2.4 bohr radii, then integrated along isotherms to obtain
values of $\Delta G_{\text{tr}}$ which were then compared with
CCI-computed values. The results were found to agree within the
relevant error bars. Since the CCI method is more computationally
efficient we applied it for the computation of the final, eight
$k$-point results.  We also estimated the quantum correction to the
classical free energy resulting from the fluctuations around the
classical trajectories of the nuclei. Using the first term of a
Wigner-Kirkwood expansion in $\hbar$\footnote{E. L. Pollock, personal
  communication}, we estimate the free energy correction at 5000~K and
200~GPa to be of the order of 0.01 eV per atom or less, and can
consequently be neglected.

\begin{figure}[htb]
        \includegraphics[width=8.5cm]{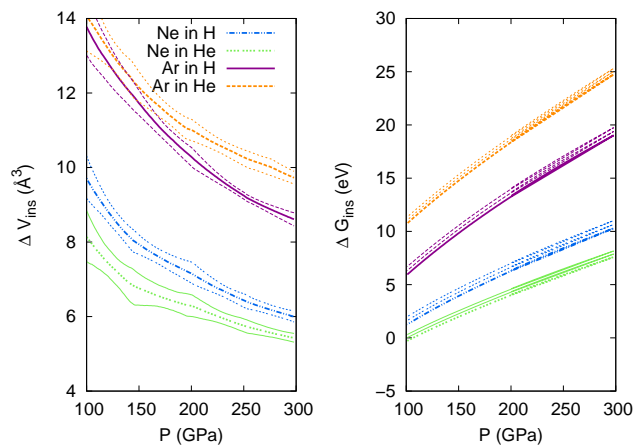}
        \caption{At left, the difference in volume $\Delta V_{\text{ins}}$ between
          the pure hydrogen/helium cells and the cells with a single Ne/Ar atom added isobarically at
          5000~K. At right, the computed difference in free energy $\Delta G_{\text{ins}}$ between the pure H/He cells and the cells with a single Ne/Ar atom added for pressures between 100 and
          300 GPa at 5000~K.}
\label{dvdg}
\end{figure}

\begin{table}
\begin{tabular} {c c| c c|c c}
\hline
T&P & \multicolumn{2}{c}{ $\Delta G_{\text{Tr}}$ (eV), CCI} & \multicolumn{2}{c} { $\Delta G_{\text{Tr}}$ (eV), IM} \\
(K)&(GPa) & Ne & Ar & Ne & Ar \\
\hline
200 & 3000 & -2.45(33) &  4.88(33) & -2.01(37) & -0.10(31) \\
200 & 5000 & -2.36(46) &  5.08(45) & -1.17(41) & 0.41(36)  \\
200 & 7000 & -2.42(63) &  5.59(66) & -1.14(48) & -0.38(43) \\
\hline
100 & 5000 & -1.61(47) & 4.73(47) & $--$ & $--$ \\
300 & 5000 & -2.78(46) & 5.65(46) &$--$  &$--$ \\
\hline
\end{tabular}
\caption{Computed CCI $\Delta G_{\text{Tr}}$ values for neon and argon
  in pure hydrogen vs pure helium phases as a function of temperature and
  pressure. A negative sign indicates preference for solubility in the
  helium phase. Also shown are the $\Delta G_{\text{Tr}}$ values computed using the ideal mixing (IM) approximation, that is, with $G = U + PV - TS_{\text{ideal}}$ where $S_{\text{ideal}}$ is the entropy of the ideal gas. The resulting $G_{\text{Tr}}$ values are significantly smaller than those computed with the CCI method.}
\label{table1}
\end{table}

The computed values of $\Delta G_{\text{Tr}}$ for neon and argon are
given in Table \ref{table1}. For neon at 200~GPa we find $\Delta
G_{\text{Tr}}$ values of approximately $-2.4$~eV with temperature
variation from 3000~K to 7000~K producing only a small variation in
$\Delta G_{\text{Tr}}$. The negative sign here indicates a preference
for solubility in helium. Thermodynamic integration of the
pressure-volume curves as shown in Figure 2(a) from 200~GPa shows
that the helium preference becomes more pronounced with increasing
pressure. For argon, we see contrasting behaviour: at 200 GPa and
5000K we have a $\Delta G_{\text{Tr}}$ of $+5.1 \pm 0.45$ eV with the
positive sign now indicating a preference for the hydrogen phase. The
magnitude of the preference for hydrogen solubility increases somewhat
with both temperature and pressure.

Following the work of Roulston and Stevenson \cite{roulston-eos-95},
we expect the rate at which neon is removed from the upper envelope to
be related to the loss rate of helium by

\begin{equation}
\frac{ {dX_{\text{Ne}}}}{dt} = X_{\text{Ne}} \exp \left( \frac{\Delta G_{\text{Tr}}}{k_B T}\right) \frac{d X_{\text{He}}}{dt}.
\end{equation}

\begin{figure}[htb]
        \includegraphics[width=7.5cm]{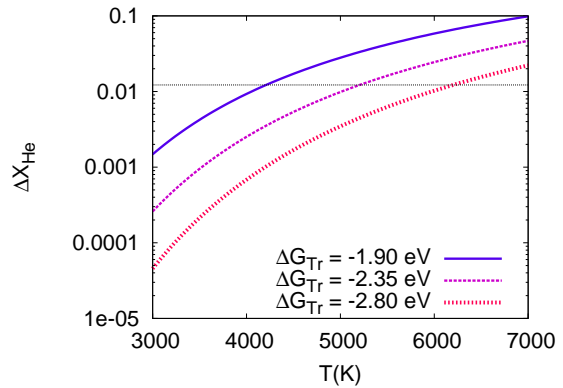}
        \caption{Relationship between the temperature of the immiscibility region and the change in helium concentration which would be required to produce the observed 90\% depletion of neon, for  $\Delta G_{\text {Tr}}$ values of -2.35 $\pm$ 0.45 eV. The observed value of 1.2\% for $\Delta X_{\text{He}}$ is marked with a line.}
\label{tvsdx}
\end{figure}

This implies that the observed depletion of neon will be approximately
given by

\begin{equation}
\log \left( \frac{X_{\text{Ne}}^1}{X_{\text{Ne}}^0} \right) = \left( X_{\text{He}}^0 - X_{\text{He}}^1 \right) \exp \left( \frac{\Delta G_{\text{Tr}}} {k_{\text B} T} \right),
\end{equation}

where $X_{Q,0}$ and $X_{Q,1}$ refer to the original (protosolar) and
present-day molar atomic concentrations of species $Q$ in the upper
Jovian atmosphere, respectively. Based on the measurements of Von Zahn
\emph{et al.} \cite{vonzahn-jgr-98} for the current helium
concentration and the estimate of Lodders \cite{lodders} for the
protosolar concentration, we find a helium depletion $X_{\text{He}}^0 -
X_{\text{He}}^1$ value of approximately 1.2\%. Combining this with
values of $\Delta G_{\text{Tr}}$ of $-2.35$ $\pm$ 0.45 eV  for
neon partitioning, we obtain the
relationship between $T$ and neon depletion shown in
Fig.~\ref{tvsdx}. The observed value of approximately 0.1 for the neon
depletion ratio corresponds to $T$ values of between approximately
4000~K and 6000~K. This is consistent with the expected temperature of the
immsicibility region. The computed value of $\Delta G_{\text{Tr}}$ is
thus consistent with the assumption that the observed depletions of
both helium and neon are due entirely to helium rain within the
hydrogen-helium immiscibility layer.

For argon, the positive value of $\Delta G_{\text{Tr}}$ implies that
Ar will be almost entirely excluded from the He phase. Since the
helium phase remains only a very small portion of the planet this will
lead only to a miniscule enhancement in the argon content of the upper
atmosphere. This implies that the measured concentration,
approximately $1.6$ times the solar value \cite{niemann-science-96},
should be close to the true argon concentration of the planet as a
whole.

\begin{figure}[htb]
        \includegraphics[height=6.5cm,angle=270]{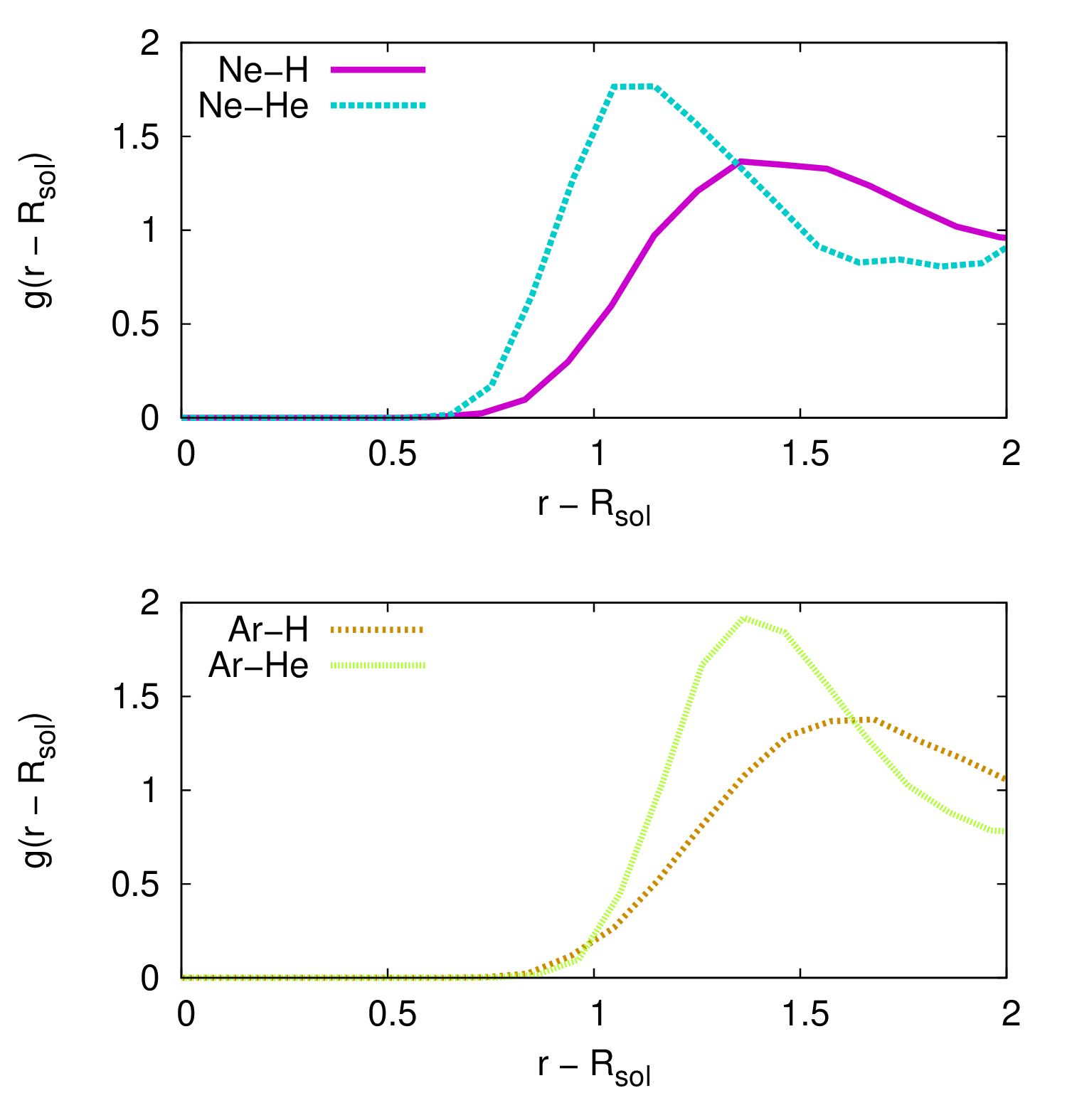}
        \caption{Pair correlation functions g($r - R_{\text{sol}})$ for distances between solvent (H,He) and solute (Ne,Ar) atoms at 200~GPa and 5000~K. The curves are shifted by the effective radius $R_{\text{sol}}$ of the solvent atom  in each case, determined from the point where the solvent-solvent $g(r)$ crosses 0.5. $R_{\text{sol}}$ is 0.37~$\text{\AA}$ for hydrogen and 0.58~$\text{\AA}$ for helium.}
\label{gr}
\end{figure}

The difference in solubility behaviour between neon and argon invites
further examination. The difference in the free energies of insertion
is governed primarily by the volume change $\Delta V_{\text{ins}}$
associated with the insertion at constant pressure of the noble gas atom
into the pure-solvent cell. As shown in Fig.~2(a) the effective volume
of neon is larger by 0.86~$\text{\AA}^3$ at 200~GPa and 5000~K in hydrogen than
in helium, while argon shows the opposite trend, being larger by
0.73~$\text{\AA}^3$ in helium than hydrogen. In Fig.~\ref{gr} we plot the
pair correlation function $g(r)$ for the solvent atoms surrounding
each species of noble gas atom. The $g(r)$ curves have been shifted by
$R_{\text{sol}}$, the effective radius of the solute atom
derived from the point at which the $g(r)$ for H-H or He-He crosses
0.5. There is a clear difference in the exclusion behaviour of neon
and argon, with the small-distance tail of the H-Ar curve allowing a
closer effective approach than in helium, in contrast to of the H-Ne
and He-Ne curves where helium approaches more closely.

As a possible interpretation, we note that in the P,T
range of interest the H atom is essentially ionized \cite{ion1,ion2} while the He atom
retains its electrons. A helium atom thus is repelled from the Ne/Ar
atom by the electron-electron interactions dominated by Pauli
exclusion, whereas hydrogen atoms may more easily penetrate the outer
shells and are repelled primarily by core-core repulsion. The Ar
atom's additional electron shell thus gives it a larger effective
volume to exclude the helium atom, but much less so for the hydrogen.
If this model is correct then we would expect the noble gases krypton
and xenon to likewise exhibit a preference for hydrogen solubility, a
behaviour consistent with their apparent observed lack of depletion in
the upper atmosphere \cite{niemann-science-96}.

In this work we have considered only pure helium and pure hydrogen
phases. In practice the helium phase will have very little hydrogen,
but the hydrogen-dominant phase still contains some helium
\cite{morales-pnas-09}, however we do not expect this to qualitatively
change the results. Another limitation not considered in this study is
whether the partitioning coefficient changes as the neon concentration
in helium increases; it should be noted that the required molar
concentration of neon in the pure-helium phase will be quite large. We
also cannot exclude, based on this study, the possibility that neon
forms its own phase, however we consider this unlikely due
to the small initial Ne concentration.

These results strongly support the existence of hydrogen-helium phase
separation in Jupiter as an explanation for the observed Ne
depletion. We have also shown that argon will be preferentially
excluded from a helium droplets explaining the observed lack of
depletion of this element. Further work to more accurately determine
the location of the hydrogen-helium immiscibility line at low pressure
and low temperature would allow us to make a quantitative estimate of
the neon concentration in Saturn to be tested by future
missions. Furthermore, neon may be added as a tracer in laboratory
experiments to detect the phase separation of H-He mixtures because
neon scatters X-rays more strongly.

{\bf Acknowledgements:} This work was supported by NASA and
NSF. Computational resources were provided in part by NERSC and
NCCS. We thank D. Stevenson for discussions and M. Morales for
providing source code.

\bibliographystyle{h-physrev3}

\newpage

\end{document}